\documentclass{aa}  
\newcommand{\ngc}[1]{NGC~{#1}}

\usepackage{graphicx}
\usepackage{natbib}
\usepackage{subfig}

\usepackage{txfonts}

\usepackage[colorlinks=true,
    linkcolor=blue, citecolor=blue, filecolor=blue, urlcolor=blue]{hyperref}

\newcommand{\msol}{\text{M}_\odot}
\newcommand{\lsol}{\text{L}_\odot}
\newcommand{\rsol}{\text{R}_\odot}

\newcommand{\teff}{$T_{\text{eff}}$}

\begin{document}

   \title{Discovery of a double white dwarf in the Galactic globular cluster NGC 6397}

   \author{Fabian Göttgens \inst{1}
          \and Marilyn Latour \inst{1}
          \and Ulrich Heber \inst{2}
          \and Sebastian Kamann \inst{3}
          \and Kyle Kremer \inst{4}
          \and Sven Martens \inst{1}          
          \and Stefan Dreizler \inst{1}
          \
          }

   \institute{Institut für Astrophysik und Geophysik, Georg-August-Universität Göttingen, Friedrich-Hund-Platz 1, 37077 Göttingen, Germany\\  \email{fabian.goettgens@uni-goettingen.de}
   \and
    Dr.~Karl~Remeis-Observatory \& ECAP, Astronomical Institute, Friedrich-Alexander University Erlangen-Nuremberg, Sternwartstr.~7, 96049 Bamberg, Germany
   \and
   Astrophysics Research Institute, Liverpool John Moores University, IC2 Liverpool Science Park, 146 Brownlow Hill, Liverpool L3 5RF, United Kingdom
   \and
   Department of Astronomy \& Astrophysics, University of California, San Diego, La Jolla, CA 92093, USA
   }

   \date{Received MONTH DD, YYYY; accepted MONTH DD, YYYY}

  \abstract
   {Binaries in the cores of globular clusters are known to prevent the gravitational collapse of the cluster, and simulations predict that the core of NGC 6397 contains a large number of white dwarfs (WDs), of which many are expected to be part of a binary system.
   
   In this work, we report the discovery of a compact binary system consisting of two WDs in the centre of the Galactic globular cluster NGC~6397. 
The system, known in the literature as NF1, was observed as part of a MUSE radial-velocity survey aiming at characterizing the binary population in the centre of NGC 6397. 
The spectral analysis of NF1 provides an effective temperature of 16\,000~K and a surface gravity (log $g$) of 5.72~(cgs), which is consistent with an extremely low-mass He-core WD nature. This is further supported by the mass of $0.23 \pm 0.03~\msol{}$ obtained from fitting the star's spectral energy distribution using its HST magnitude in various filters. 
The system has a circular orbit with a period of 0.54 days.
The radial velocities show a large semi-amplitude of 200 km/s, implying a minimum mass of $0.78~\msol{}$ for the invisible companion, which is likely another WD, or a neutron star if the inclination of the system is smaller than about 50$^\circ$. 
Some significant residuals in radial velocity remain with our best orbital solution and we tested whether a model with a third body can explain these deviations. While this possibility seems promising, additional measurements are needed to confirm whether the star is actually part of a triple system.
   } 


   \maketitle

\section{Introduction}
%\object{[CGC98] NF 1}
%\object{NGC 6397}
The dense cores of globular clusters (GCs) contain an abundance of stellar exotica, including the massive remnants of stars, such as white dwarfs (WDs), neutron stars, and black holes. 
Binaries influence the dynamical state of their host GC. As the GC evolves, the more massive objects, including binaries, black holes, and massive white WDs ($>$ 0.8 $\text{M}_\odot$), accumulate in the GC centre due to mass segregation. 
In the GC core, the binaries support the GC against core collapse. With time, they are either destroyed or ejected from the GC after interactions with other stars \citep{kremer_role_2019}.
In particular, very compact binaries consisting of heavy objects emit potentially detectable amounts of gravitational waves \citep{kremer_white_2021}.

Extremely-low mass WDs (ELMs) are He-core WD with a mass below $0.3~\msol{}$. Because of their low mass and their degenerate interior, ELMs are the largest and brightest WDs.
They are known to exist in the Milky Way, typically as part of a compact binary system \citep[see, e.~g.][]{webbink_evolution_1975,marsh_low-mass_1995,brown_elm_2020}. 
This is intrinsically linked to their formation,
which can only be explained as a result of binary interaction since the universe is too young to form low-mass WDs from single star evolution. 
In GCs, ELMs were so far observed as companions of millisecond pulsars \citep[e.~g.][]{matasanchez_psr_2020, cadelano_psr_2020} and recently discovered as companions of blue straggler stars in NGC~362 \citep{dattatrey_globules_2023}. 
Previous photometric observations of bright WDs and their central concentration in \ngc{6397} \citep{cool_cataclysmic_1998} suggest that they could be part of binary systems with a more massive but invisible companion, most likely a carbon-oxygen WD \citep{hansen_helium_2003, strickler_helium-core_2009}. 
Their position in the colour-magnitude diagram \citep{strickler_helium-core_2009} and the spectroscopic analysis of one of these bright WDs \citep{edmonds_cataclysmic_1999} are consistent with them being He-core WD.

\ngc{6397} is a core-collapsed GC situated at a distance of 2.4~kpc, making it the second-closest GC to Earth. 
Kinematical analyses of \ngc{6397} suggest that its core contains either a dark component, e.~g. an intermediate-mass black hole or many stellar remnants \citep{kamann_muse_2016}, such as WDs \citep{arnold_direct_2025}.
Monte-Carlo simulations of the cluster also suggest that it contains a central, diffuse cluster of WDs in its core \citep{kremer_white_2021,vitral_stellar_2022}. 
This subsystem has a mass of about $10^3~\msol{}$ and consists mostly of carbon-oxygen and oxygen-neon WDs. The simulations also contain a low number of He-core WDs which scales with the primordial binary fraction because He-core WDs only form in binary systems. 
In this central cluster of WDs, binaries consisting of two WDs are also formed due to the high stellar density. These binaries are hardened and are either ultimately destroyed when the WDs merge after a gravitational-wave inspiral, or ejected from the GC after interaction with single stars or binaries \citep{kremer_white_2021, weatherford_stellar_2023}.
These processes are hard to observe since most WDs are intrinsically faint, unless they are part of a cataclysmic variable where the WD accretes from a non-degenerate companion, i.e. a main-sequence star.
WD binaries consisting of at least one He-core WD are much easier to detect from radial velocity measurements than those without a He-core WD because these are brighter than the other types, and since they are also less massive, their RV amplitude is larger. 

NF1 in NGC~6397 is part of a group of UV-bright stars located close to the centre of the cluster \citep[][see Fig.~\ref{fig:finding_chart} for a finding chart]{cool_cataclysmic_1998}. 
These stars are located bluewards of the main-sequence and below the main-sequence turn-off in colour-magnitude diagrams.
UV light curves show that some of these stars are photometrically variable on timescales of hours and  ``flicker'' on even shorter timescales, while others show no photometric variability (``nonflickerers'', NF). 
While the flickerers are known or suspected cataclysmic variables, \citet{cool_cataclysmic_1998} propose that the NFs are low-mass He-core WD in a binary, either with a faint and low-mass main-sequence star ($m < 0.15~\msol{}$), a cool WD, or a neutron star.
\citet{hansen_helium_2003} use theoretical cooling models and relations between orbital parameters and exchange times in star clusters to infer that the binary companions of the NFs are more likely to be CO-WDs than neutron stars, and that they were likely formed in the last $10^8$~years.
Recently, \citet{pichardo_marcano_second_2025} find that the UV light curve of NF1 varies with a low amplitude, which they attribute to spots caused by magnetic fields. However, they do not detect a clear periodicity in their signal.

In this paper, we use NF1~AB to designate the binary system with its more massive, invisible, primary component NF1~A and its visible secondary NF1~B.
The paper is structured as follows: Section~\ref{sect:data} lists the MUSE and Hubble Space Telescope (HST) data used in this work, Section~\ref{sect:analysis} presents the methods used for the spectroscopic and spectral energy distribution (SED) fits, and for the radial velocity analysis. Section~\ref{sect:results} present the stellar and orbital parameters, and we discuss our results in Section~\ref{sect:discussion_prime}. Finally, we conclude in Section~\ref{sect:conclusion}.

\section{Data}
\label{sect:data}

We use spectra obtained with MUSE, an optical integral-field spectrograph at the VLT \citep{bacon_muse_2010}. 
MUSE has a large field-of-view of $1'\times 1'$ with spatial sampling of 0.2$\arcsec{}$ and is equipped with adaptive optics (AO). 
It covers the spectral range from 4750 to 9350~\AA{} at a constant sampling of 1.25~\AA{} with a spectral resolution of 1800 in the blue and 3500 in the red.
The data were obtained as part of an observation programme to detect binary systems in the centre of NGC~6397 (Programme ID 0111.D-2117(A), PI: S. Dreizler). A total exposure time of 4.4~h was split into 20 epochs, which were observed from May 2023 to August 2023 using AO (see Table~\ref{tab:observation_dates}).
Each MUSE datacube consists of four exposures with different derotator positions, each with an exposure time of 200~s.
We reduced the data with the MUSE pipeline \citep[][version 2.8.9]{weilbacher_data_2020}. 
The data cubes produced by the pipeline were then analysed with \texttt{PampelMuse} \citep{kamann_resolving_2013}, which extracts spectra of individual stars using point spread function fitting.
The individual spectra of NF1~AB have a signal-to-noise (S/N) ranging from 7 to 18, depending on the observing conditions. 
We did not include archival MUSE data from 2014 \citep{kamann_muse_2016,husser_muse_2016} and 2017 because they were taken without AO and have a lower S/N. 

We used the HST Legacy Survey of Galactic Globular Clusters \citep[HUGS, ][]{piotto_hubble_2015,nardiello_hubble_2018} and the Hubble Space Telescope Atlases of Cluster Kinematics \citep[HACKS, ][]{libralato_hubble_2022} photometric catalogues of \ngc{6397} to collect the photometric measurements of NF1~AB from the UV to the red visual band (see Table~\ref{tab:mags}). In addition, the HACKS catalogue also contains relative proper motions that we use to verify the cluster membership of NF1~AB.

\section{Analysis methods}
\label{sect:analysis}

\begin{figure*}
\centering
   \includegraphics[width=\textwidth]{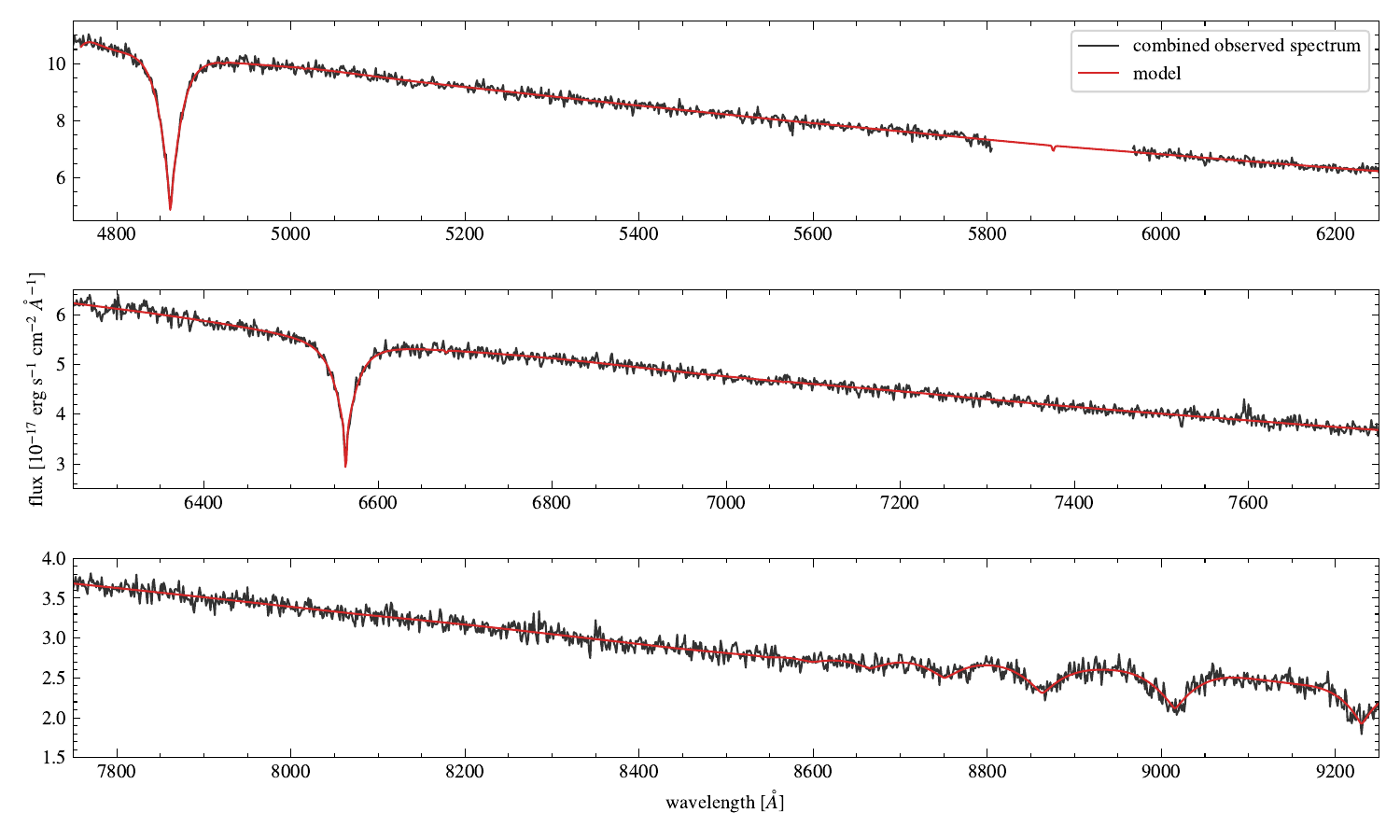}
     \caption{Best fit of the combined MUSE spectrum of NF1~AB. The observed spectrum is combined from 19 single spectra with a total exposure time of 4.2 hours.}
     \label{fig:spectrum}
\end{figure*}

\subsection{Spectral fit and spectral energy distribution}

The MUSE spectra of NF1~AB were first fitted using \texttt{spexxy} and the Göttingen Spectral Library \citep{husser_new_2013, husser_muse_2016} to estimate their radial velocities and reproduce the telluric lines. The telluric contamination was then removed from the individual spectra, they were shifted to rest-frame velocity, and finally co-added to obtain a higher S/N spectrum (see Fig.~\ref{fig:spectrum}).

The Göttingen Spectral Library does not cover the parameter range of NF1~AB, so we used a different model grid to fit its spectrum.
We used models computed with the ADS approach \citep{przybilla_testing_2011,irrgang_proto-helium_2021} which uses the ATLAS12, DETAIL, and SURFACE codes to produce hybrid LTE/NLTE model atmospheres suitable for compact hot stars such as those on the blue horizontal branch and the hottest blue stragglers. 
These models have been used by \citet[see their Sect. 3 for a detailed summary of the models and the parameter coverage of the grids]{latour_shotglas_2023} to analyse horizontal branch stars and some hot blue stragglers in NGC~6752 and $\omega$ Centauri.
The model grids cover four dimensions: \teff, log~$g$, helium abundance, and metallicity. Most importantly, they cover the appropriate atmospheric parameter range of NF1~AB\footnote{We note that available model grids for hydrogen-rich (DA) WD only cover log~$g$ values larger than 6.5 or 7.0 (e.g., \citealt{koester_white_2010}).} (see Sect.~\ref{sect:sed_fit_results}). 
He-core WDs are less compact than the typical CO-core WDs, but gravitational settling is nevertheless effective at removing metals from their photosphere and they are expected to have a low metallicity (see, e.g., \citealt{gianninas2014}). 
Therefore and because the stars in NGC~6397 have a metallicity of $\text{[Fe/H]}=-2.0$ (\citealt[2010 edition]{harris_catalog_1996}), we used our lowest solar-scaled metallicity ($\text{[M/H]}=-2.0$) grid to fit the spectrum of NF1~AB.  
The spectrum was fitted using the Interactive Spectral Interpretation System (ISIS) with an updated version of the $\chi^2$-minimization method presented in \citet{irrgang_new_2014}. In the case of NF1~AB, the only visible spectral lines are the hydrogen Balmer lines H$_\alpha$ and H$_\beta$ as well as a few Paschen lines. The fit of the MUSE spectrum allows us to constrain \teff{}, log~$g$, and estimate an upper limit to the He abundance.

Besides fitting the MUSE spectrum to derive the atmospheric parameters, the SED of NF1~AB was also fitted to derive additional stellar parameters, namely the radius, luminosity, and mass of the star.
The synthetic fluxes in the various HST filters are constructed from the grid of ATLAS12 model atmospheres.
The observed SED of NF1~AB is defined from the magnitude measurements listed in Table \ref{tab:mags}. A general description of the SED fitting method is presented in \citet{heber_spectral_2018} and \cite{irrgang_quantitative_2018}. For NF1~AB, we make use of the known distance and reddening of NGC~6397 and fix these values to $D=2.4 \pm 0.1$~kpc and $E(B-V)=0.18$~mag (\citealt{brown_high-precision_2018,maiz_apellaniz_validation_2021,harris_catalog_1996}, 2010 edition).  
Since the surface gravity cannot be well constrained by photometry, it is also kept fixed to the value obtained from the MUSE spectrum. The two parameters left free to vary during the SED fit are then \teff{} and the angular diameter ($\Theta = R/D$). From $\Theta$, we directly obtain the radius $R$ since the distance is known. From the radius, the luminosity and the mass are obtained via the formulae 
$L=4\pi R^2 \sigma T^4 _{\rm eff}$ and $M=gR^2/G$.

\subsection{Radial velocities and orbital modelling}

\begin{figure*}
\centering
   \includegraphics[width=\textwidth]{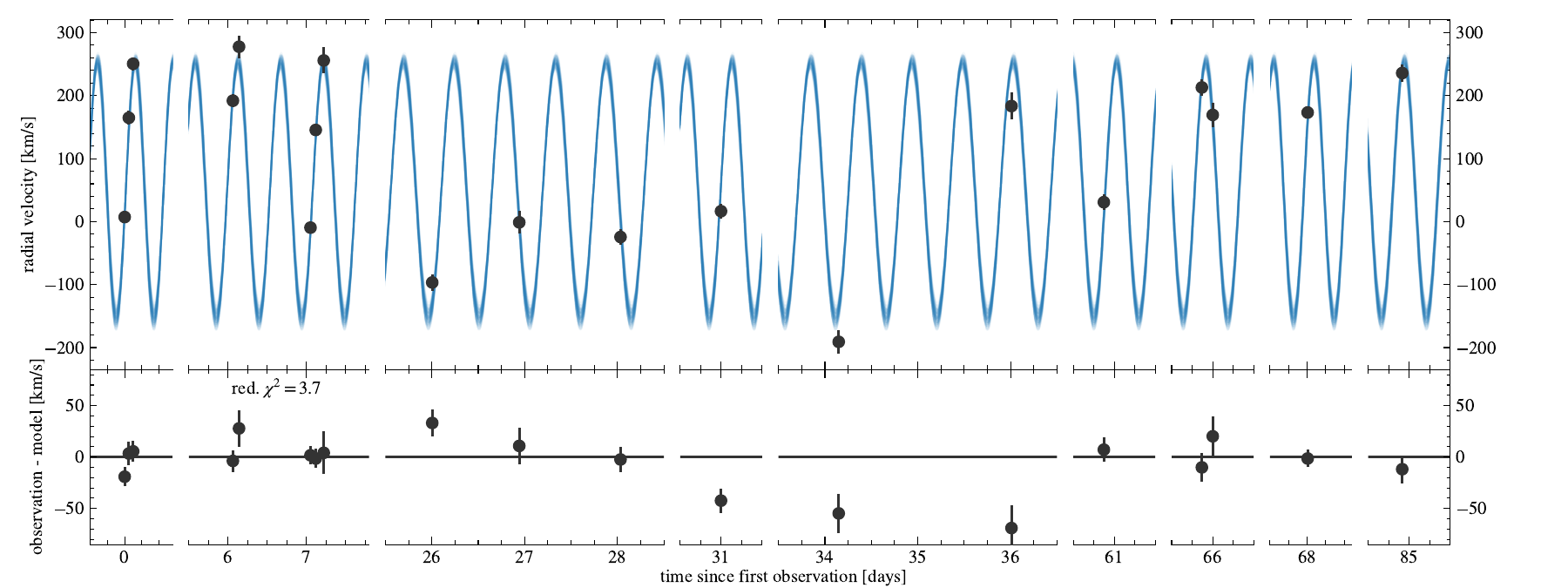}
     \caption{Radial velocities of NF1~B observed in 2023 and predicted radial velocity curves computed from 100 parameter sets randomly drawn from the posterior. The residuals are computed relative to the median prediction.
     The first observation started on JD 2460082.68297265 (May 18 2023 4:23 UTC).}
     \label{fig:rvs}
\end{figure*}

The initial radial velocities derived from the MUSE observations and the Göttingen Spectral Library show a clear amplitude variation, indicating that the star is part of a binary system. Because the spectral library does not contain WD models, we re-computed the radial velocities of the individual spectra using this time the best-fitting synthetic spectrum from the grid described above.
For each observed spectrum, we fit the radial velocity and the parameters of the telluric components using \texttt{spexxy} and the best-fit synthetic spectrum to minimize the squared residuals between the model and observations. 
Figure~\ref{fig:rvs} shows the radial velocities obtained this way. 
Table~\ref{tab:observation_dates} lists the observations dates, the radial velocities as well as their uncertainties, and the estimated $S/N$ values.

In order to estimate the orbital parameters, we use an orbit model together with the nested sampling code \texttt{ultranest} \citep{buchner_ultranest_2021}.
The free parameters of the Kepler model are the systemic velocity $\gamma$, the semi-amplitude $K$, the period $p$, the eccentricity $e$, the argument of periapse $\omega$ and the mean anomaly $\phi$ at $t_0$, where $t_0$ is the time of the first observation. 

We define our likelihood function $L$ as
\begin{equation}
\log L = - \frac{1}{2} \sum_i \frac{[v_i - v(t_i)]^2}{\sigma^2_{v,i}}, 
\label{eq:likelihood}
\end{equation}
where $v_i$ and $\sigma^2_{v,i}$ are the observed radial velocities and their respective uncertainties, and $v(t_i)$ is the predicted velocity given the orbital parameters.

We can not directly compute the companion mass because the inclination of the system is unknown. 
For the Kepler case, a minimum companion mass ($m_2 \sin i$) can be numerically estimated from the orbital parameters using \citet{murray_keplerian_2010}
\begin{equation}
    m_2 \sin i = K \left[ \frac{p (m_1 + m_2)^2}{2 \pi G} \right]^{1/3} \sqrt{(1-e^2)}.
\label{eq:mass}
\end{equation}
Table~\ref{tab:priors} lists the priors used in the nested sampling analysis. 
In addition to a flat prior for the systemic velocity $\gamma$, we also compute solutions with a normal distribution centred on the cluster mean velocity with a standard deviation of 5~km/s and a normal distribution centred on the mean observed radial velocity of NF1 B with a standard deviation of 20~km/s. 
The lognormal distribution used as period prior is based on results from N-body models in GCs from \citet[][$N_c=10^5$ case]{ivanova_evolution_2005}.

\section{Results}
\label{sect:results} 

\begin{table}
    \centering
    \small
    \renewcommand{\arraystretch}{1.25}
    \caption{Properties of the binary system NF1 AB.}
    \begin{tabular}{ll}
        \hline
        \hline
        \multicolumn{2}{c}{NF1 A (primary)} \\
        stellar type & white dwarf \\
        min. mass & $0.78\pm0.06~\msol{}$ \\
        \\
        \multicolumn{2}{c}{NF1 B (secondary)} \\
        stellar type & He-core white dwarf \\
        $T_\text{eff}$ & $16 140^{+170}_{-120}$~K \\
        log~$g$ & $5.72^{+0.04}_{-0.03}$ (cgs) \\
        radius & $0.109 \pm 0.005~\rsol$ \\
        luminosity & $0.74 \pm 0.07~\lsol$ \\
        mass & $0.23\pm 0.03~\msol$ \\
        \\
        \multicolumn{2}{c}{Orbital parameters} \\ 
        period $p$ & $0.54351 \pm 0.00003$~d \\
        semi-amplitude $K$ & $202 \pm 6$~km/s  \\
        eccentricity $e$ & $<0.03$ \\
        systemic velocity $\gamma$ & $50.1 \pm 4.1$~km/s \\
        argument of periastron $\omega$ & $1.7^{+2.5}_{-1.1}$ \\
        $t_0$ (JD) & 2460082.6941611907 \\
        mean anomaly at $t_0$& $3.2^{+1.2}_{-1.5}$ \\
        \\
        \multicolumn{2}{c}{Sampling statistics} \\
        evidence $\log Z$ &  $-50.0 \pm 0.2$ \\
        KL divergence $H$ & $24.5 \pm 0.1$ \\
        max. $\log L$ & $-23$ \\
        \hline        
    \end{tabular}
    \label{tab:stellar_orbit_parameters}
\end{table}

\begin{figure}
    \centering
    \includegraphics[width=0.45\textwidth]{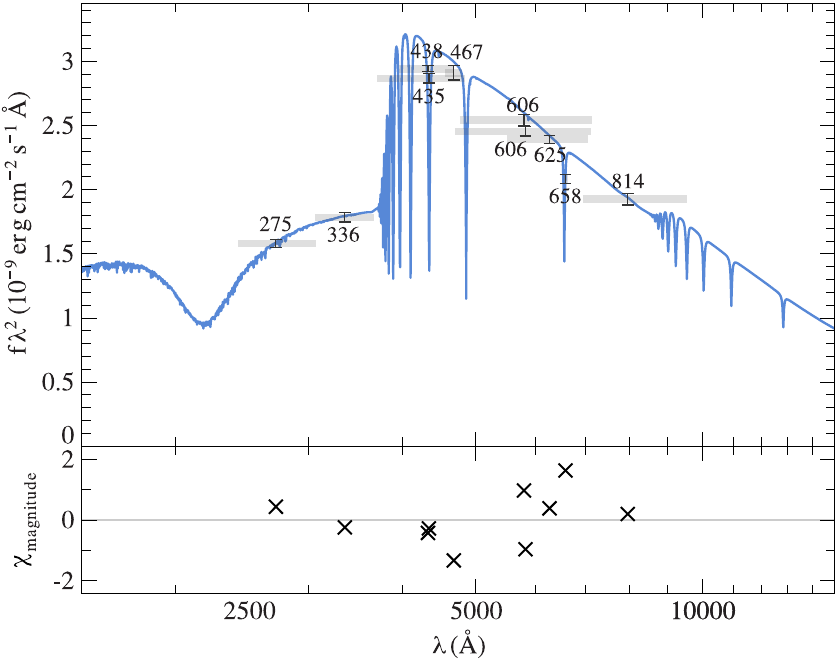}
    \caption{SED fit of NF1~B. On the y-axis, we plot the flux $f_{\lambda}$ multiplied by $\lambda^2$. The best-fit model is shown in grey and the observed magnitudes in the various filters are indicated along with their central wavelength. The wavelength coverage of each filter is indicated with gray horizontal bars. The bottom panel shows the uncertainty-weighted residuals $\chi$ = (mag$_{\rm model}$-mag$_{\rm observed}$) / uncertainty. }
    \label{fig:sed}
\end{figure}

\subsection{Cluster membership}
First, we verify that NF1~AB is a member of NGC~6397.
We use the proper motions published in the HACKS dataset \citep{libralato_hubble_2022}. Its values are
$\mu_\alpha \cos \delta = 0.080 \pm 0.076$~mas/yr and $\mu_\delta = 0.118 \pm 0.122$~mas/yr relative to the mean cluster motion.
These values are well inside the bulk of the GC stars (see Fig.~\ref{fig:pm}) and thus indicate that NF1~AB is part of \ngc{6397}.
In principle, the systemic velocity of the binary system can also be used to verify the stars' membership to the GC, but we leave the discussion of this aspect to Sect.~\ref{discussion:residuals}.

\subsection{Atmospheric and stellar parameters}
\label{sect:sed_fit_results}
The best fit of the MUSE spectrum with the synthetic WD spectra is illustrated in Fig.~\ref{fig:spectrum}. We derive \teff{} = 16140 $^{+170}_{-120}$ K, $\log g$ = 5.72 $^{+0.04}_{-0.03}$~(cgs), and $\log (N_{\rm{He}}/N_{\rm tot}) = -2.58^{+0.21}_{-1.6}$. The absence of He lines in the spectrum and the large uncertainty on the helium abundance derived essentially indicate an upper limit of $\log (N_{\rm{He}}/N_{\rm tot}) \lesssim -2.5$.
Using the surface gravity and He abundance obtained from the spectral fit, we then proceed with the SED fit. The resulting best fit is illustrated in Fig.~\ref{fig:sed}. From the SED we obtain \teff{}$^{\rm SED}=16220 \pm 160$~K and $R=0.109 \pm 0.005~\rsol$. We note here that the effective temperature derived from the SED is in perfect agreement with the spectroscopic value. From the given radius, we also derive $L = 0.74 \pm 0.07~\lsol$ and finally $M = 0.23 \pm 0.03~\msol$. 

These parameters are consistent with an ELM WD nature, confirming that NF1~B, the visible component of the binary NF1~AB, is a low-mass He-core WD. However, the star has a surface gravity lower than the majority of ELMs at this \teff{}, which have log~$g >$ 6.0 \citep{brown_elm_2020, gianninas2014}. Most of the low-gravity ELMs (log~$g <$ 6.0) are also referred to as proto-WD, meaning that their temperature is still increasing and they have not yet entered their final cooling phase (see Sect. 2 of \citealt{istrate_models_2016} and the evolutionary tracks displayed in Fig.~\ref{fig:logg_teff_zoom}).

\subsection{Comparison with evolutionary models}
\label{sect:nf1_evol_models}

\begin{figure}
    \centering
  \includegraphics[width=0.5\textwidth]{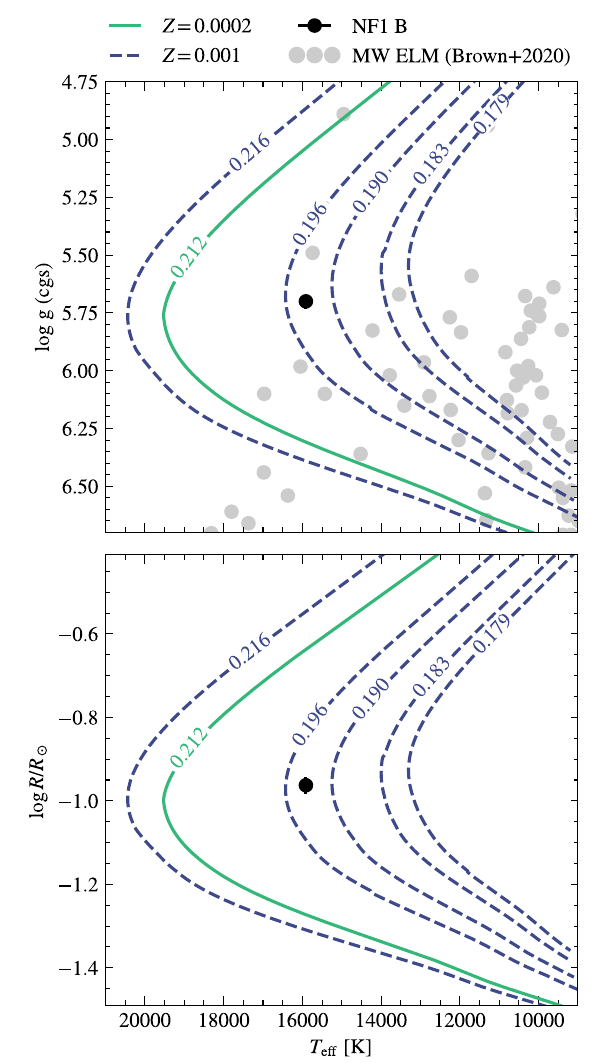}
  \caption{Position of NF1~B in the log~$g$$-$$T_\text{eff}$ diagram (top) and log~$R$$-$$T_\text{eff}$ diagram (bottom) compared to selected model tracks including rotation and diffusion from \citet{istrate_models_2016}. The labels along the tracks give the stellar mass in solar masses.
  The grey points are Milky Way ELMs from the survey of \citet{brown_elm_2020}. }
  \label{fig:logg_teff_zoom}
\end{figure}

The mass of NF1~B can also be estimated by comparing its atmospheric parameters, $T_\text{eff}$ and log~$g$, to He-core WD evolutionary models of different masses and metallicities. 
We use the evolutionary tracks of \citet{istrate_models_2016} which predict log~g and $T_\text{eff}$ during the evolution of low-mass He-core WDs at different metallicities. These models take into account the effects of elemental diffusion and rotational mixing. 
NGC~6397 has $\text{[Fe/H]}=-2.0$ \citep[][2010 edition]{harris_catalog_1996}, corresponding to $Z=0.00014$, which is comparable to the lowest metallicity ($Z=0.0002$) considered in the models of \citet{istrate_models_2016}.

Figure~\ref{fig:logg_teff_zoom} shows the position of NF1~B in the log~g-$T_\text{eff}$ and radius-\teff\ diagrams along with evolutionary tracks for the final cooling phase of He-core WDs. 
Among the tracks at $Z=0.0002$, only the one with the lowest mass ($0.212~\msol{}$) goes through the parameter space represented in Fig.~\ref{fig:logg_teff_zoom} but it predicts a higher \teff\ at both the log~$g$ and radius values of NF1~B. To see evolutionary sequences at lower \teff, we need to use the models computed with $Z=0.001$. Although they have different metallicities, the tracks behave in a smooth way as the mass decreases. Based on these tracks, we would expect the mass of NF1~B to be close to $0.2~\msol{}$, which is in good agreement with the value estimated from the SED fit and distance to the cluster. We note that the position of the cooling tracks is also affected by the mass of the hydrogen envelope remaining on the ELM WD. The thickness of the hydrogen envelope depends on whether the mass transfer was stable or unstable. A thick H-envelope is formed in the former process  while a thin envelope is predicted by the latter one \citep[see, e.g.][]{althaus_2025}. This introduces an additional uncertainty in the mass determination from evolutionary tracks, which we estimate to be similar to that derived from the SED fit. 

The star is located very close to the turn off point of the evolutionary tracks, meaning close to the maximum \teff\ reached by the models, and its position is consistent both in terms of log~$g$ and radius.
We note that this agreement was not guaranteed, because log~$g$ is obtained from the spectral fit while the radius is obtained from the angular diameter which do not strongly depend on the surface gravity.

\subsection{Orbital parameters}

Using \texttt{ultranest}, the likelihood defined in Eq.~\ref{eq:likelihood} with a Kepler orbit model, the priors in Table~\ref{tab:priors} including a flat prior on the systemic velocity, and the observed radial velocities, we find a unimodal solution with a period of $0.54351 \pm 0.00003$~d, a systemic velocity $\gamma=50.1 \pm 4.1$~km/s, a very low eccentricity $e<0.03$, and a semi-amplitude $K=202\pm 6$~km/s (see Table \ref{tab:stellar_orbit_parameters}). 
The distributions of these parameters and more are plotted in Fig.~\ref{fig:corner} for different choices of the prior for the systemic velocity $\gamma$.
We draw 100 samples from the posterior distribution obtained with the flat prior for the systemic velocity and calculate the predicted radial velocity curve for each sample.
Figure~\ref{fig:rvs} shows the individual radial velocity measurements and the 100 predicted radial velocity curves.
The phase-folded data are shown in Fig.~\ref{fig:phase_rv}.
The residuals have a reduced $\chi^2 = 3.7,$ indicating that the nominal radial velocity uncertainties do not account for the full variation observed in the velocities. 
The reported uncertainty in the period is very low, which is not unusual given the short period and the long observation time span (see e.g. \citealt{brown_elm_2020}).

With the orbital parameters from the posterior distribution and the mass from the visible He-core WD (see Sections~\ref{sect:sed_fit_results} and \ref{sect:nf1_evol_models}), we estimate the mass of the unseen star using Eq.~\ref{eq:mass}.
Assuming an edge-on orbit ($\sin i = 1$), we derive a minimum mass of $0.78\pm0.06~\msol{}$ for NF1~A when assuming a mass of $0.23 \pm 0.03~\msol{}$ for the He-core WD (NF1~B).
For the other two prior choices for $\gamma$, we find a minimum mass of $0.74 \pm 0.06 ~\msol{}$ in case of the prior centred on the mean observed radial velocity and $0.83 \pm 0.06~\msol{}$ with the prior centred on the mean cluster radial velocity. 

\section{Discussion}
\label{sect:discussion_prime}

\begin{figure}
\centering
   \includegraphics[width=0.5\textwidth]{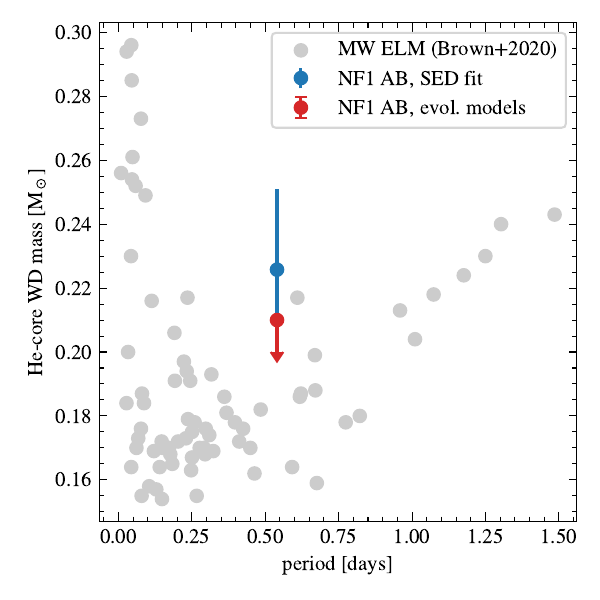}
     \caption{Mass of NF1~B versus its orbital period compared to those of ELM binaries in the Galactic field \citep{brown_elm_2020}. }
     \label{fig:brown_mw_mass_period}
\end{figure}

\subsection{Data quality}
Given the large residuals between the observed RVs and those predicted by our binary solution, we want to make sure that this is not due to issues with the data, e.g. contamination by close sources or data reduction problems.

If contamination occurs due to crowding, the observed spectrum of a star is then a combination of one or multiple other sources and the stellar parameters derived from a contaminated spectrum are not representative of the observed star. 
Due to the varying observing conditions, the influence of any potential contamination would also vary and thus cause changes in the apparent brightness.
We check for contamination by calculating magnitudes in the Bessel I, R, and V bands for each observed spectrum of NF1~AB. 
We measure the variation by calculating the standard deviation of the 19 observations in each band.
NF1~AB has standard deviations of 0.19, 0.14, and 0.13~mag in the Bessel I, R, and V bands, respectively. 
In Fig.~\ref{fig:brightness_variation}, we compare these values to the median standard deviation of 187 stars with similar F606W~magnitude ($\pm 0.25$~mag) as NF1~AB. 
These values are $0.15^{+0.17}_{-0.04}$ mag, $0.16^{+0.19}_{-0.04}$ mag, and $0.18^{+0.26}_{-0.06}$ mag, for the same three bands, where the uncertainties represent the $1\sigma$ range of values obtained from the 187 comparison stars. 
We can conclude from these values that NF1~AB does not vary more than similarly bright stars.
To check the reliability of our radial velocities, we choose close-by ($\le 5\arcsec$) comparison stars with a similar brightness ($\pm 0.25$~mag in F606W) as NF1~AB. 
The finding chart (Fig.~\ref{fig:finding_chart}) shows our five selected stars relative to NF1~AB (red circle). The reduced $\chi^2$ values of their radial velocities are 0.8, 1.0, 1.0, 1.1, and 1.4, see Fig.~\ref{fig:phase_rv} (bottom). We thus find that these stars do not show any significant variation in their radial velocities and conclude that the RVs we observe for NF1~B and their variation are neither caused by contamination due to close-by stars nor by data reduction issues.

\subsection{Formation}
ELM systems in the Galactic field follow two distinct branches in the period-mass plane \citep{brown_elm_2020}, as shown in Fig.~\ref{fig:brown_mw_mass_period}. 
One branch comprises systems with short periods and a wide range of secondary (the He-core WD) masses, which were likely formed following a common-envelope phase.
A second branch contains systems with orbital periods that seem to increase proportionally with the He WD mass. This branch is likely the result of Roche-lobe overflow \citep{li_formation_2019, brown_elm_2020}.

Given the mass of NF1~B, we show in Fig.~\ref{fig:brown_mw_mass_period} that the position of the system is consistent with the branch extending to longer periods, and therefore was likely formed by Roche-lobe overflow. 
The system is however positioned slightly above this branch, especially when the mass derived from the SED fit is used.
Observations of ELM in the Milky Way \citep{brown_elm_2016} and simulations of ELM formation \citep{li_formation_2019} usually find or assume circular orbits, as we also obtained for NF1~AB.

\subsection{Nature of the invisible component}
The invisible component has a minimum mass of $0.82\pm0.06~\msol{}$.
In order to estimate whether a low-luminosity main-sequence star could be the companion, we use a \textsc{parsec} \citep{bressan_span_2012} isochrone that fits the CMD of NGC~6397.
According to this isochrone, a main-sequence star with a mass of $0.6~\msol{}$ has a F606W magnitude of about 18.6 mag and would already be slightly brighter than NF1~AB (see Table~\ref{tab:mags}). 
Therefore, any main-sequence companion would need to have a mass lower than $0.6~\msol{}$, but this is ruled out by our orbital solution. 
Possible stellar types of the companion are thus a WD or a neutron star.
When we sample values of $\sin i$ to compute a distribution of inclination-corrected companion masses, we find that in about 35 to 40~\% of all cases the companion has a mass above $1.45~\msol{}$ ($i \lesssim 50^{\circ}$), consistent with a neutron star.
The closest known radio sources to NF1~AB are likely not associated with the system because they are located between $17\arcsec$ and $20\arcsec$ away \citep[ATCA1, ATCA2, ATCA3 in][]{tudor_maveric_2022}.

\subsection{Can a triple explain the large residuals?}
\label{discussion:residuals}
As Fig.~\ref{fig:rvs} shows, our best orbital solution produces large residuals, starting at 26 days after the first observation, of about $+50$~km/s that gradually decrease until day 36 where it is about $-50$~km/s. 
These residuals cause a high reduced $\chi^2$ of 3.7 for the overall fit, much larger than the expected value of about 1.
Another peculiarity of the best orbital solution is the large systemic velocity ranging from $37.1 \pm 3.1$~km/s to $52.4 \pm 4$~km/s, depending on its prior. Given the mean cluster radial velocity of $18.51 \pm 0.08$~km/s \citep{baumgardt_catalogue_2018}, the system has a relative radial velocity of at least $19\pm3$~km/s and proper motions of about 1 and 3~km/s, which in total is close to but below the central escape velocity of 21.5~km/s \citep{baumgardt_catalogue_2018}. 
A possible explanation for the large residuals and the unusual systemic velocity is the existence of a third component. 
We perform a fit using the N-body integration code \texttt{rebound} \citep{rein_rebound_2012, rein_ias15_2015} for a triple star system and our measured RV data. 
We find a bimodal solutions for all three prior choices of the systemic velocity and all solutions have a lower reduced $\chi^2$ than the binary solution.
For the flat prior and the normal prior centred on the mean of the observed RVs, we find either a third component of $0.8~\msol$ with a period of 40~days or a more massive star with a broad mass distribution ($m_3 > 0.8~\msol{}$) and a broad period distribution of more than 60~days. In both cases, the eccentricity of the third component is moderate to high ($e > 0.4$).
For the normal prior centred on the average RV of the cluster, we find either a third component with a period of 20 days, low eccentricity and a mass of $0.8~\msol{}$ or a more massive star ($m_3 > 1.4~\msol{}$) with a period of 60 days and low to moderate eccentricity ($e < 0.4$).
Whether the systemic velocity $\gamma$ of the triple solution is in better agreement with the mean GC velocity also depends on the choice of the prior. For a normal prior centred on the mean of the observed RVs and a dispersion of 5~km/s,
we find $\gamma \approx 60$~km/s and thus a worse agreement with the GC systemic velocity.  For a normal prior centred on the average RV of the cluster, we obtain $\gamma \approx 25$~km/s and the use of a flat prior results in $\gamma \approx 50$~km/s.
The use of a flat prior in the binary fit also results in a systemic velocity of about 50~km/s.

Simulations of globular clusters predict triple systems with both configurations: hierarchical triple systems with two WDs as the inner binary and either a WD or a neutron star as the outer companion. In these simulations, triples with a WD as the third component occur ten times more often than those with a neutron star \citep{fragione_demographics_2020}.
However, the triple model likely overfits the data given that we have 19 observed RVs and the triple model includes 12 free parameters. 
To test this, we generate artificial RV data from our binary solution and add as noise the permuted residuals from the original binary fit. This generated data set can also be convincingly fitted with a triple star configuration. 
Although the triple star model has its advantages, we currently do not have enough data to confirm that there is third component in the NF1~AB system.

\section{Conclusion}
\label{sect:conclusion}

We presented the spectral analysis of NF1~AB, a binary containing a UV-bright ELM WD in NGC~6397.
By combining 19 MUSE spectra to determine the spectroscopic properties of the visible component, we found an effective temperature of $16 140^{+170}_{-120}$~K, a log~g of $5.72^{+0.04}_{-0.03}$~(cgs) and a helium abundance below $\log (N_{\rm{He}}/N_{\rm tot}) = -2.5$. Our fit of the star's SED, based on HST photometry and the known distance and reddening of the cluster, confirmed the \teff{} obtained from the spectra and allowed us to derive
a stellar mass of $0.23 \pm 0.03~\msol{}$,
These properties are in good agreement with evolutionary model tracks of He-core WDs, confirming what was suspected from previous analyses based solely on photometry \citep{cool_cataclysmic_1998,strickler_helium-core_2009}. 
The RV curve shows large variations with a semi-amplitude of about 200~km/s, confirming that NF1 is a binary system. 
After fitting it with a Kepler model, we determine that the orbit is essentially circular with a period of 0.5435~days. 
The unseen component, NF1~A, has a minimum mass of $0.78\pm0.06~\msol{}$ and is thus heavier than the visible component. Since it does not contribute to the flux or the spectrum of the system, it is a compact stellar remnant, most likely another WD or a neutron star if the inclination of the system is $\lesssim 50 ^{\circ}$. 
NF1~AB appears to be very similar to the ellipsoidal variable star V46 in NGC~6121 \citep{kaluzny_1997,kaluzny_2013}. The visible component of V46 has very similar atmospheric parameters as NF1~B, implying an equally low mass, but a shorter orbital period of 0.0872 days, thus explaining the presence of ellipsoidal variations caused by its compact companion \citep{otoole2006}.

Galactic ELM binary systems follow two distinct branches in the mass-period plane \citep[e.g.][]{brown_elm_2020}.
By plotting NF1~AB in this plane, we find that it falls onto the branch that most likely results from Roche-lobe overflow instead of a common-envelope phase.

Our best fit for the orbital solution reproduced well the observed RVs, although some large ($\pm$50 km/s) residuals remained between day 26 and 36 (after the first observation). Thus, we tested whether a triple-system solution can reduce those residuals.
We found that a third star with a period of either 20 or 80~days could explain the observed deviations, while the properties of the inner binary remain unchanged within their uncertainties.
However, given the larger number of free parameters in the triple-star model and the limited number of RV measurements, additional observations are needed to confirm the presence of a third body in this system.

The discovery of a double WD sheds light onto the hidden population of faint WDs in globular clusters.
A comprehensive survey of WDs and other remnants in binaries will enable us to further test the predictions of globular cluster evolution, in particular the large number of WDs in binaries predicted in the core of this cluster.

\begin{acknowledgements}
We thank the anonymous referee for their constructive comments.
Based on observations collected at the European Organisation for Astronomical Research in the Southern Hemisphere under ESO programme(s) 0111.D-2117(A). F.G. and M.L. acknowledge funding from the Deutsche Forschungsgemeinschaft (grants DR 281/41-1, LA 4383/4-1). S.M acknowledges funding from the Federal Ministry of Research and Technology (grant 05A23MGA).
SKA gratefully acknowledges funding from UKRI through a Future Leaders Fellowship (grants MR/T022868/1, MR/Y034147/1).
We thank A. Irrgang and M. Dorsch for developing and maintaining the analysis tools used and S. Hämmerich for providing the grid of model atmospheres and synthetic spectra. This research has made use of NASA’s Astrophysics Data System Bibliographic Services.

\end{acknowledgements}
\bibliographystyle{aa}
\bibliography{aa58594-25}

\begin{appendix}
\section{Additional material}

\begin{table}[h]
    \centering
    \renewcommand{\arraystretch}{1.15}
    \caption{Observation start dates and radial velocities for the individual MUSE spectra ($\Delta$BJD is the Barycentric Julian date $d-2460082$~days).}
    \begin{tabular}{rrrr}
        \hline
        $\Delta$BJD [days] & $v$ [km/s] & $\sigma_v$ [km/s] & $S/N$ \\
        \hline
0.69416 &     7.2 &   9.4 &   17.1 \\
0.75493 &     164.8 & 11.4 &  13.0 \\
0.81813 &     250.6 & 10.4 &  13.9 \\
6.75674 &      191.9 & 10.6 &  15.0 \\
6.83526 &      277.6 & 17.8 &  9.1 \\
7.74736 &      -9.6 &  8.9 &   18.2 \\
7.81454 &      145.4 & 9.4 &   16.0 \\
7.91642 &      255.8 & 20.8 &  8.7 \\
26.69952 &      -96.7 & 12.8 &  12.8 \\
27.63684 &      -1.2 &  17.9 &  10.2 \\
28.72792 &       -24.5 & 12.2 &  13.8 \\
31.69250 &      16.5 &  11.9 &  14.6 \\
34.84307 &       -190.9 & 18.8 &  9.1 \\
36.70474 &      183.4 & 21.8 &  7.3 \\
61.56664 &      30.8 &  12.0 &  13.6 \\
66.56286 &      212.9 & 13.7 &  12.9 \\
66.69583 &      169.4 & 19.2 &  8.5 \\
68.70052 &      173.2 & 8.6 &   17.8 \\
85.60798 &      235.8 & 13.5 &  10.4 \\
        \hline
    \end{tabular}
    \label{tab:observation_dates}
\end{table}

\begin{table}[h]
    \caption{HST-derived photometric magnitudes of NF1~AB. }
    \centering
    \renewcommand{\arraystretch}{1.15}
    \begin{tabular}{lrrl}
        \hline
        filter name & magnitude & uncertainty & reference \\
        \hline
        WFC3-F275W & 18.107 & 0.015 &  (1) \\
        WFC3-F336W & 18.277 & 0.0088 & (2), (3) \\
        WFC3-F438W & 19.0698 & 0.0034 & (2), (3) \\
        WFC3-F467M & 19.0809 & 0.0005 & (2), (3) \\
        WFC3-F606W & 18.9699 & 0.0077 & (2), (3) \\

        ACS-WFC-F435W & 19.0554 & 0.0049 & (1) \\
        ACS-WFC-F606W & 19.0027 & 0.0087 & (1) \\
        ACS-WFC-F625W & 18.9652 & 0.0058 & (1) \\
        ACS-WFC-F658N & 18.914 & 0.0121 & (1) \\ 
        ACS-WFC-F814W & 18.9389 & 0.0225 & (1) \\
        \hline
    \end{tabular}
    \tablebib{
(1)~\citet{libralato_hubble_2022}; (2)~\citet{soto_hubble_2017}; (3)~\citet{nardiello_hubble_2018}.
}
    \label{tab:mags}
\end{table}

\begin{figure}[p]
    \centering
    \includegraphics[width=0.45\textwidth]{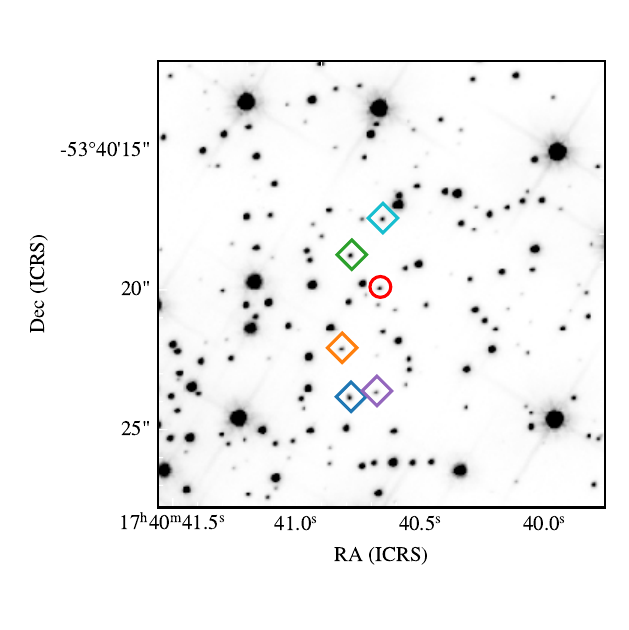}
    \caption{Finding chart \citep[ACS-WFC F606W,][]{sarajedini_acs_2007,anderson_acs_2008} of NF1~AB (red circle) in the centre of NGC~6397. All other marked stars are comparison stars (diamonds). The region shown is about 15 arcsec by 15 arcsec in size.}
    \label{fig:finding_chart}
\end{figure}

\begin{table*}[p]
    \centering
    \small
    \renewcommand{\arraystretch}{1.15}
    \caption{Priors used in the binary orbit modelling.}
    \begin{tabular}{cccl}
        \hline
        Parameter & Prior & Unit & Description \\
        \hline
        $\gamma$ & flat & km s$^{-1}$ & systemic velocity \\
        $p$ & lognormal($\mu=0$,\, $\sigma=2$) & $e^{2}$ days & period \\
        $e$ & triangular($a=0$, $b=0.95$, $c=0$) & 1 & eccentricity \\
        $K$ & triangular($a=0$, $b=500$, $c=0$) &  km s$^{-1}$ & semi-amplitude \\
        $\omega$ & uniform(0, $2\pi$) & rad & argument of periapse \\
        $\phi$ & uniform(0, $2\pi$) & rad & mean anomaly at $t_0$ \\
        \hline
    \end{tabular}
    \label{tab:priors}
\end{table*}

\begin{figure}[p]
  \centering
  \includegraphics[width=0.45\textwidth]{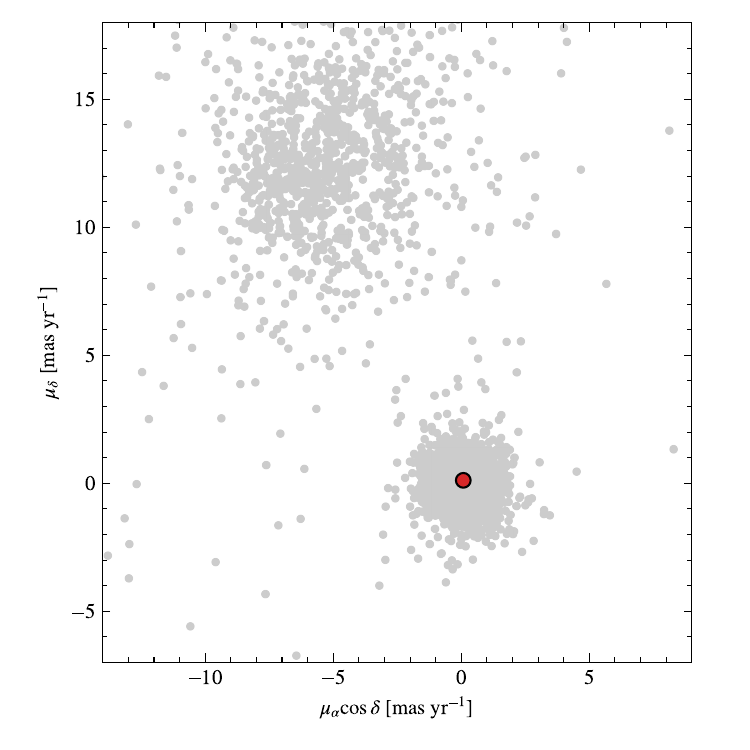}
  \caption{Relative proper motions of stars within the HST field of view centred on NGC~6397 \citep{libralato_hubble_2022}. The overdensity located at (0, 0) consists of GC members, the red dot is NF1~AB.}
  \label{fig:pm}
\end{figure}

\begin{figure}[p]
 \centering
 \includegraphics[width=0.49\textwidth]{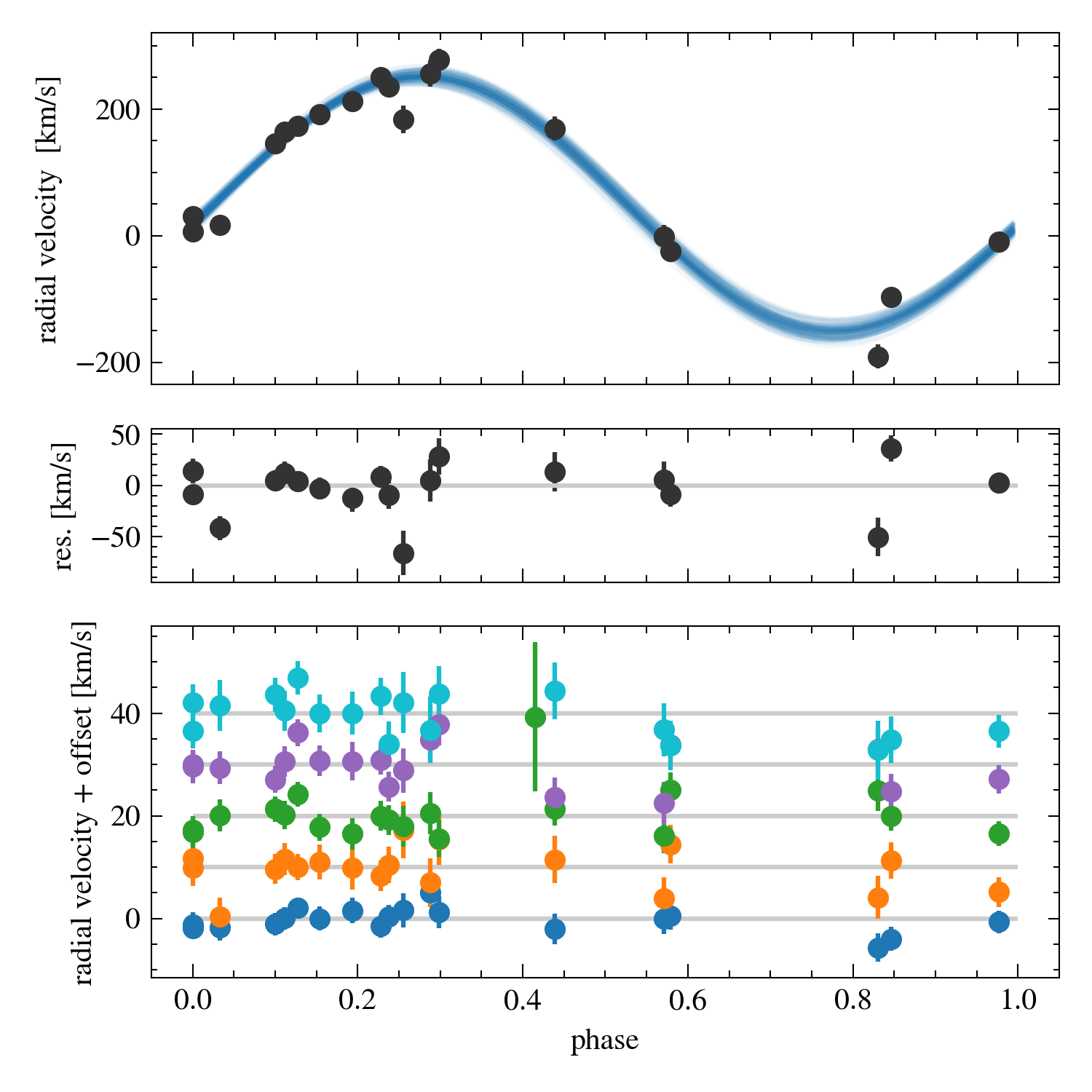}
 \caption{Phase-folded RV curve for NF1~AB (top), residuals (middle), and five close-by comparison stars (bottom, same colour-coding as Fig.~\ref{fig:finding_chart}). 
 For the RV curve of each comparison star, we add an offset and subtract the median RV. Note the different scales on the y-axis.}
  \label{fig:phase_rv}
\end{figure}

\begin{figure}[p]
 \centering
 \includegraphics[width=0.49\textwidth]{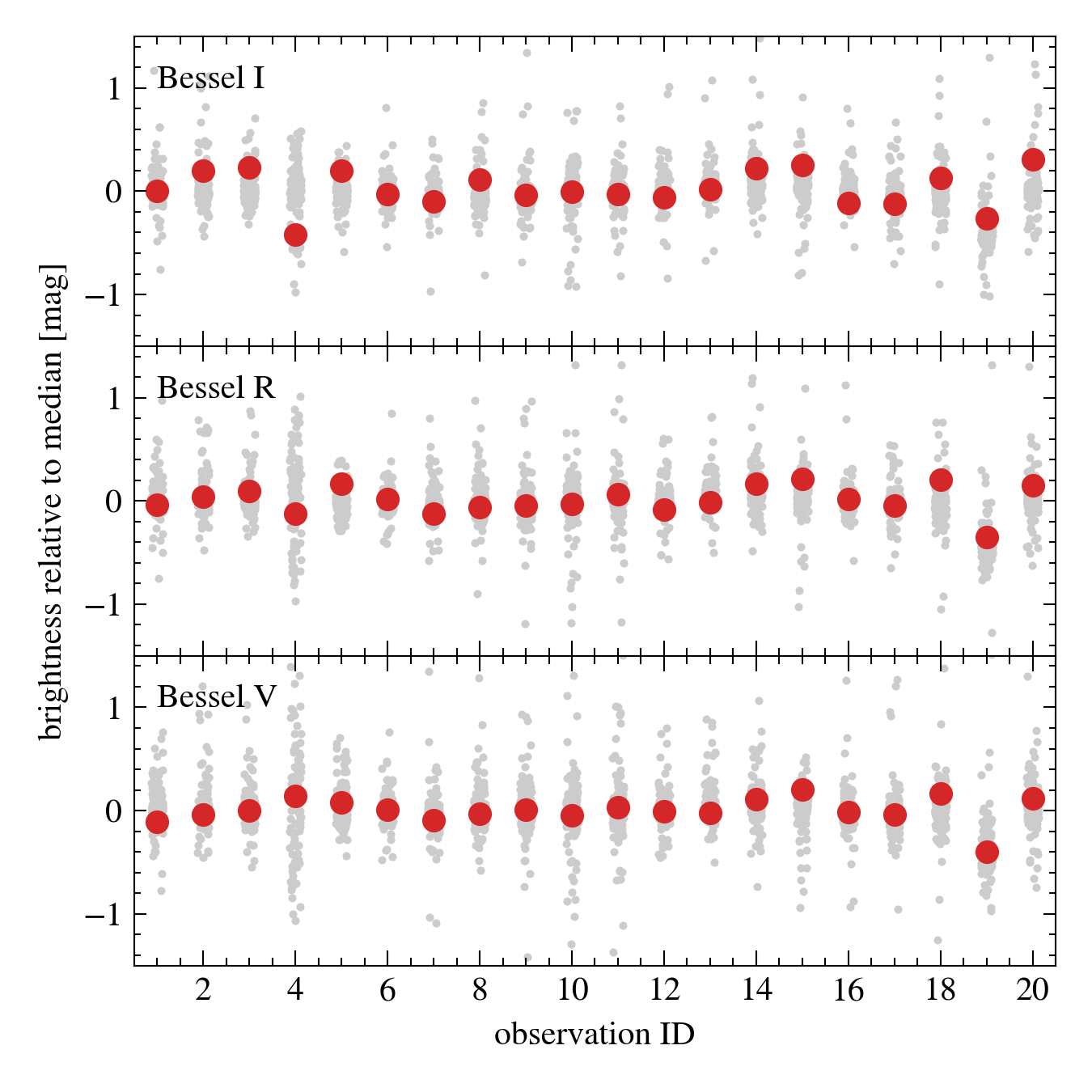}
 \caption{Brightness in three different filters computed from the MUSE spectra of NF1~AB (red circles) and 187 similarly bright stars relative to their respective median brightness as a function of observation. This plot includes the one observation which we omitted from the RV analysis due to low S/N.}
  \label{fig:brightness_variation}
\end{figure}

\begin{figure*}[p]
  \centering
  \includegraphics[width=0.999\textwidth]{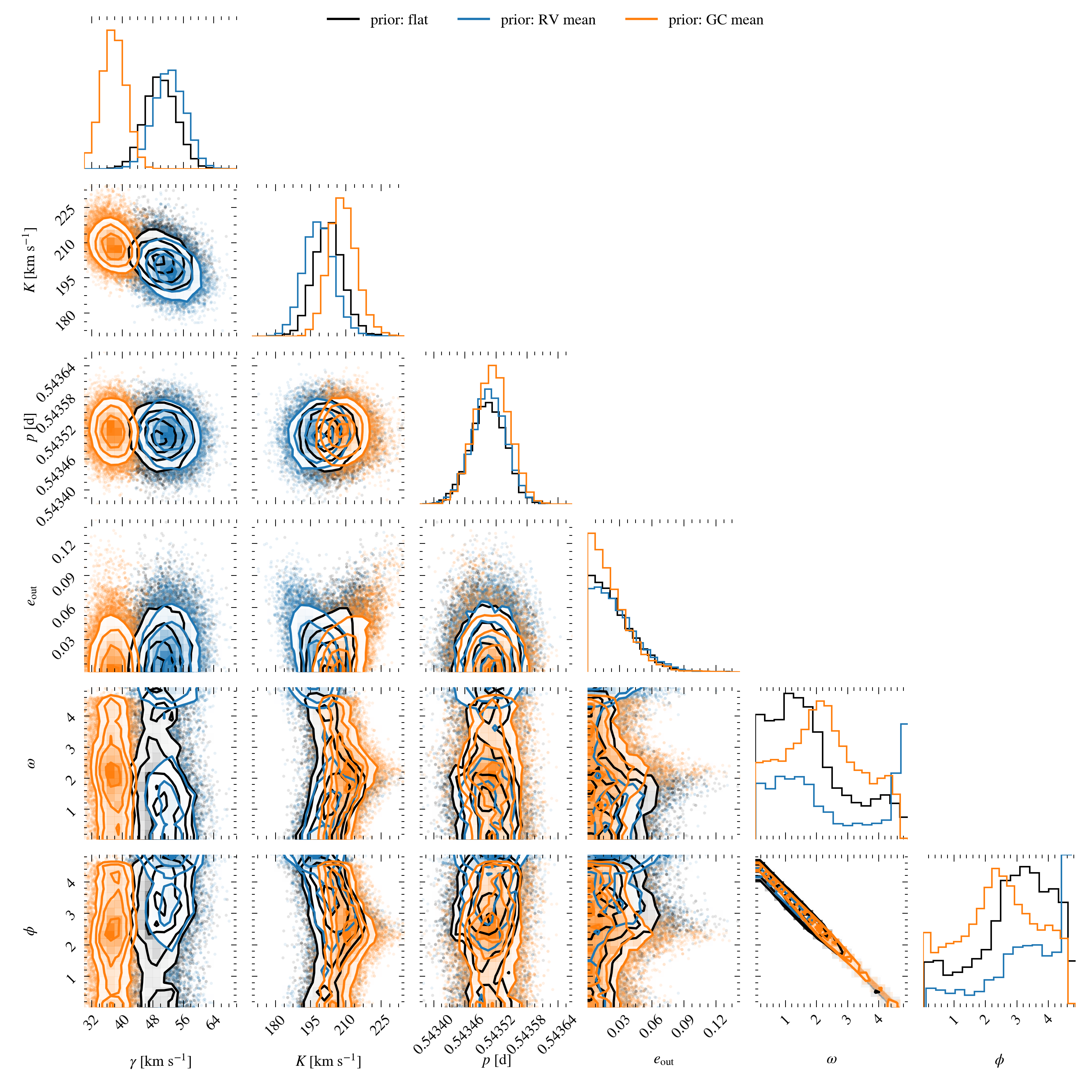}
  \caption{Corner plot of the binary orbit model for different prior choices of the systemic velocity $\gamma$. See Table~\ref{tab:priors} for a description of the parameters.}
  \label{fig:corner}
\end{figure*}

\end{appendix}

\end{document}